\documentclass[a4paper,11pt]{article}

\usepackage[colorlinks=true,linkcolor=black,citecolor=black,urlcolor=blue]{hyperref}
\usepackage{geometry}
\usepackage{fancyhdr}

\usepackage{array}
\usepackage{graphicx} 
\usepackage{multirow} 
\usepackage{dcolumn}
\newcolumntype{.}{D{.}{.}{-1}}

\usepackage{natbib}
\bibpunct{(}{)}{;}{a}{,}{,}

\usepackage{amsmath, amssymb, amsfonts, bm}
\usepackage{enumitem}
\usepackage{calc}

\usepackage[margin=30pt]{caption}

\usepackage[table]{xcolor}
\usepackage{titlesec}
\titleformat{\subsection}
            {\bfseries}{\thesection.\thesubsection.}{1em}{}

\titlespacing*{\subsection}{0pt}{1.5em}{0.2em}

\renewcommand\eqref[1]{Equation~\ref{#1}}

\renewcommand{\thesection}{\arabic{section}}
\renewcommand{\thesubsection}{\arabic{subsection}}

\setlist[2]{noitemsep}
\setenumerate{noitemsep}

\newlength{\bibitemsep}
\newlength{\bibparskip}
\let\oldthebibliography\thebibliography
\renewcommand\thebibliography[1]{%
  \oldthebibliography{#1}%
  \setlength{\parskip}{\bibitemsep}%
  \setlength{\itemsep}{\bibparskip}%
}

\fancypagestyle{firstpage} 
{
    \fancyhf{}
    \fancyhead[C]{ \includegraphics[height=2.2cm]{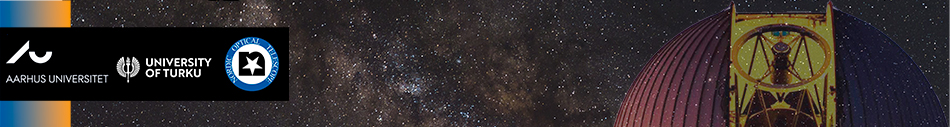}}
    \fancyfoot[C]{\thepage}
}
\fancypagestyle{allotherpages} 
{
    \fancyhf{}
    \fancyhead[R]{Nordic Optical Telescope workshop {\em NOT - a telescope for the future}, La Palma, June 2022}
    \fancyfoot[C]{\thepage}
}
\begin{document}
\pagestyle{allotherpages}
\thispagestyle{firstpage}
\vspace*{0.65cm} 

\begin{center}
{\LARGE \bf NOT Stockholm Supernovae}
\end{center}

\vspace{4pt}
\begin{center}
{\bf Jesper Sollerman}

\vspace{0.5cm}
{\small
  Department of Astronomy, Stockholm University, Sweden \\
The Oskar Klein Centre, Albanova, Stockholm University \\
Stockholm Observatory \\
}
\end{center}
\textbf{Abstract}

This proceeding contribution is a short summary of the invited talk
about observational supernova science at Stockholm University that has
been conducted at the Nordic Optical Telescope over the past 25 years,
and some expectations for the future.


\section{Introduction}

In the NOT workshop in June 2022, I delivered an invited talk about
supernova research conducted at the Nordic Optical Telescope (NOT)
from Stockholm
University. It was titled {\it The Nordic Optical Telescope and
Stockholm Supernovae 1994-2024 (my personal view)}.
This talk was given in a
semi-historical context from my own first observing runs in 1994 to
future plans for supernova classifications of LSST supernovae. The talk is
available on the NOT WWW, and here I simply give a summary
of a few of the points given in that presentation. I have been running
a supernova program at the NOT as service mode / Target-of-Opportunity for the
past 10+ years, most of them as Large Programmes. The talk was not
intended to be comprehensive, rather nostalgic.

\section{Early days}

I have been doing supernova observations for over 30 years and am a
keen user of the NOT. It was indeed the first observing runs at this
telescope that made me aim for an observational astronomy career, even though
all other supernova researchers at Stockholm observatory at that time were
doing modeling.

I remember well my first observing runs as amazingly fascinating, although
in hindsight maybe not that efficient. Filling the nitrogen dewars
yourself while integrating on a supernova, using the offline PC to find
coordinates for guide-stars to track on, screwdriver in the pocket to
fight with filter wheels. The NOT in the 1990s was a great
telescope on a fantastic site, but the instrumentation and
automation that enabled efficient observing, including service mode
and Target-of-Opportunity, came later. I argued in my talk that this
was at least a good way to learn the skills of observing, even if the
scattered epochs of supernova observations that emerged from a typical
visitor run in the 1990s is a far cry from what modern supernova
science requires. The training aspect later motivated me to run two
\href{https://ttt.astro.su.se/~solle/NOTKURS/participants.html}
{summer schools}
at the NOT with Nordic PhD students (2003 and 2006).
Other contributions in this conference proceeding focus more on the important
educational aspects of the telescope.

In my talk I also briefly reviewed the development of supernova
surveys. In the early days, a great number of nearby supernovae were
in fact discovered by dedicated amateur astronomers, and professional
astronomers were slow to take up the supernova survey strategies
developed by Fritz Zwicky in the middle of the last century. With the
High-z Supernova Search team \citep{tonry} where I participated to
search for Type Ia supernovae for cosmology in the
late 1990s, the technique was adopted with modern CCDs to image the
sky several nights apart and search for transients in the difference image
subtractions.
With subsequent pre-planned visitor runs on larger telescopes we could then
classify the newly discovered SNe. ESSENCE \citep{essence} followed a similar
scheme, whereas the SDSS-II supernova search \citep{sdssII}
employed a dedicated
telescope with a rolling search. This automatically
gave well-covered multi-band, high-cadence
supernova light curves. iPTF and its successor ZTF
\citep{graham,bellm} have added increasingly larger fields, higher
cadence and automation and are bringing transient science to the
new era that the Vera Rubin Observatory will cement.

This development and evolution in survey capacity has not only hugely
increased the number of SNe detected with time, but also the quality
of the observations. More SNe means that we find rarer objects, and there
is often a lot of physics to deduce from such outliers. Higher cadence
means we find the SNe earlier, and the untargeted search strategy
gives a less biased view of the underlying populations, which has also
revealed new classes of objects.

\section{Recent past}

I have not done the exercise to count the number of refereed papers
that have been published based on the supernova programmes at the NOT
led from Stockholm. There must be $\gtrsim100$ such papers, and deciding
which ones have been most important of course depends on who you
ask.
Some handful of PhD students have based their supernova PhD-theses on NOT
data, most of them supervised by myself on the core-collapse side of
the supernova spectrum, or by Ariel Goobar for the thermonuclear ones.
A handful of publications have made it into
high-impact journals like {\it Nature} and {\it Science} over the
years\footnote{For the 2 dozens of papers in Nature/Science journals
  that I have co-authored over the past 20 years, about a third
  include data from the NOT, including the birth of the field of Flash
spectroscopy for supernovae \citep{galyam,yaron}, and the new Type Icn
supernova class \citep{galyam22}.}. 

The reason NOT is so good for us doing transient astronomy is that it
is \href{http://www.not.iac.es/education/comics/supernovas/}
{flexible},
simple, accessible and often fast. Transient astronomy is
about getting the data when needed, and within international
collaborations we can complement or even rival larger telescopes by
being smart on when and how to get the spectra. This has allowed us to
participate in many interesting investigations, but also to take the
lead in a number of projects.

\begin{figure}[ht!]
  \centering
  \includegraphics[width=1.0\textwidth]{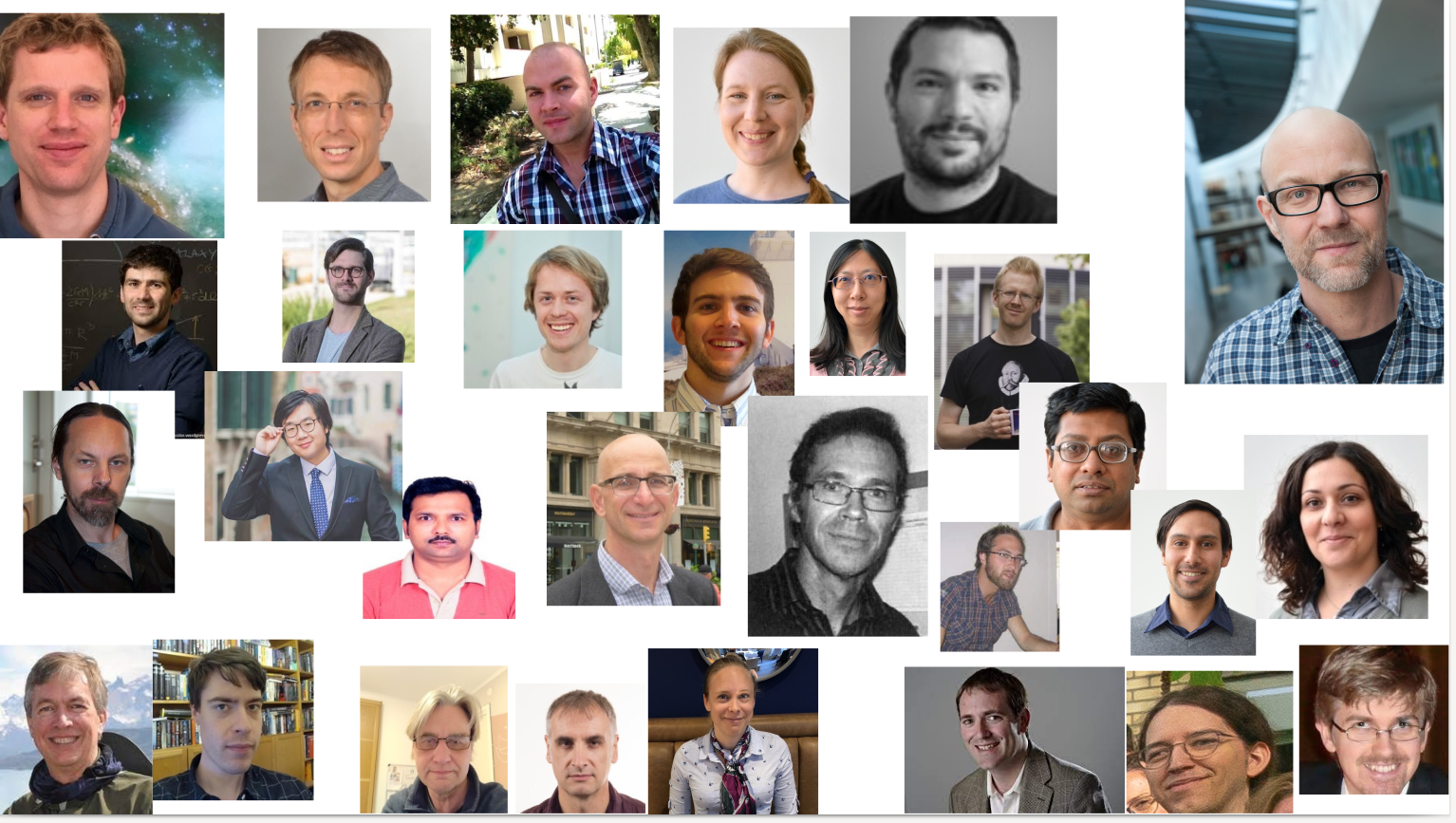}
  \caption{To pick a single slide from my presentation I would select
    this one. These are all astronomers that first-authored a refereed
    publication including observations from the NOT while they were
    working at Stockholm University. See text for further information
    on some of these articles.}
  \label{fig:Fig}
\end{figure}

Figure~\ref{fig:Fig} illustrates this by
showing a person-gallery of Stockholm scientists that have all led papers based on data from the NOT,
including work on (this is from left to right, from top to bottom in Fig.~\ref{fig:Fig}):
Helium-rich Type Ia-CSM supernova with radio detections \citep{kool},
oxygen abundances in Type II supernovae \citep{jerkstrand},
the stripped-envelope supernova iPTF13bvn \citep{fremling},
superluminous supernovae from ZTF \citep{lunnan}
the magnetar-driven iPTF15dtg \citep{taddia}, and the
enigmatic zombie-supernova iPTF14hls \citep{arcavi,sollerman}.

A stripped-envelope SN with a H-shell \citep{tartaglia},
Type Ia supernova distances \citep{johansson},
the candidate pair-instability supernova SN 2018ibb \citep{schulze2},
or hydrogen deficient Type Ibn supernovae \citep{karamehmetoglu},
and a Type Ic SN metamorphing into a Type IIn \citep{janet}, or the 
bumpy light-curve of Type IIn iPTF13z \citep{nyholm}.

The PhD-thesis papers on the now canonical Type IIb
\href{https://ttt.astro.su.se/~maer0651/sn2011dh-progenitor.html}
{SN 2011dh}
\citep{ergon2014,ergon2015},
the low-luminosity supernova SN 2020cxd \citep{yang},
the almost superluminous SN 2012aa \citep{roy},
the multiply imaged gravitationally lensed Type Ia supernova iPTF16geu \citep{goobar},
or work on circumstellar interaction supernovae like in the monumental paper on
SN 2010jl where NOT data were complemented with spectra from the
Hubble Space Telescope \citep{fransson},
or analysis of Type Ia SN spectra \citep{nordin},
the power of supernova siblings \citep{biswas},
or extinction law studies for Type Ia supernovae \citep{amanullah}, and
the full sample of stripped-envelope Type Ic SNe from PTF \citep{barbarino}.

Bumpy superluminous supernovae (SLSNe) of Type I \citep{west} and of
SLSNe Type II \citep{kangas},
constraints on hydrogen in Type Ia SNe \citep{lundqvist} and a very
detailed study of a normal Type Ia \citep{stanishev}.
The NOT spectroscopic follow-up of SDSS-II SNe Ia
\citep{ostman}
and long-lived interaction powered Type IIn supernovae
\citep{stritzinger}, an oxygen-rich stripped envelope supernova \citep{schweyer}
and finally the Crab supernova remnant and its famous pulsar \citep{sandberg}.

There is a strong legacy not only in the scientific data and these
published results, but also in the number  of researchers that have
benefited from this long-term collaboration. The relative diversity in
the topics mentioned above also illustrates that an observing
programme for transient sources needs flexibility to allow seizing the
opportunities as they arise in the universe. We must both be ready to
work on systematic and complete samples and to quickly
jump on whatever new explosion that occurs.

That Nordic transient astronomy has flourished with the help of the NOT was
also obvious from the many presentations given during this workshop.
In the distant past, the proposal I lead was the only one for supernovae at the
NOT, while today there is a plethora of projects from for example  \AA rhus, \AA
bo, Copenhagen and also several from outside the Nordic countries
(like Spain and Italy). Many of these Nordic supernova nodes are now led by
astronomers that once did an observational supernova postdoc with
us in Stockholm.
NOT has clearly been a key component for allowing astronomers in the Nordic
countries to do competitive transient science on the international scene.

\section{Nowadays}

The current main focus of my own SN research is the
\href{https://www.ztf.caltech.edu/}
{Zwicky Transient Facility} (ZTF).
With a refurbished Schmidt telescope (P48) on classical
Palomar, a super-sized 47 square-degree detector, and a very
low-resolution dedicated follow-up spectrograph (SEDM on P60), I can discover a
handful of supernovae every night, immediately get a spectrum and classify
them. Since 2018 we have trawled the Northern sky on a nightly basis and
within the Bright Transient Survey (BTS) \citep{fremlingbts,perley}
we have so far classified
over 6000 SNe. I have discovered just as many myself, which
highlights the new era we are in\footnote{Fritz Zwicky himself was for
a long time the record holder in discovering SNe, with 120 SNe
over a period of $\sim50$ years. Hard work at a cold telescope, daytime
blinking of the plates and endurance that contrasts with my own
clicking on the computer over breakfast coffee.}.
The light curves of these supernovae are directly made available to the public
via the
\href{https://sites.astro.caltech.edu/ztf/bts/explorer.php}
{Bright Transient Survey explorer} and the classification spectra are
immediately uploaded to the Transient Name Server (TNS).

In contrast to the early days of supernova science, this rolling,
untargeted and systematic system more or less automatically provides three
($gri$) photometric band light curves for supernovae and other
transients. The new era thus combines the survey strategies once
developed in the supernova cosmology framework with the rapid
follow-up capacities inherited from the fast-shooting $\gamma$-ray
burst community.

An illustrative example of how the scope of the transient astronomy
has expanded could be to think of supernova 1987A, which was the
main topic of research for the supernova group in Stockholm when I
started my PhD.
It got the designation `A' since it was the first
supernova discovered in that year, on February 24 1987. This means
that the first 54 days that year did not have a single supernova. 
In contrast, this year, 2022, 340 classified SNe had already been reported to the TNS up to February 24 (about 6 per day), so a SN discovered that day got a more complicated name such as SN 2022dek.
Of these 340 SNe, ZTF follows some 68\% and discovered around 20\%.
But the classified supernovae are only a small fraction of the
reported transients; there were in fact over 3000 transients reported
to TNS in the first 55 days of the year 2022, so only 11\% were
classified supernovae. One third of the classifications were done with
the SEDM on the P60, for which I am the main responsible.

The name of the game is now instead to recognize which of all the new
objects are interesting enough for a dedicated follow-up campaign.  We
have seldom used the NOT for pure classifications, most objects are
simply not interesting enough to waste valuable NOT time. Today the
SEDM on the P60 is the classification work horse, and we trigger NOT
mainly for objects that we already know are going to be of interest
(for some specific publication). Once I have allocated a target to
SEDM, the robotic execution is automatic, the reductions are automatic
\citep{rigault} and if it is a typical SN Ia it will automatically be
classified \citep{SNIascore} and uploaded to the TNS.
Once NOT confirms the interest of a particular object, many
other, often larger, telescopes are typically triggered for further
investigations. This is a good food-chain for transient astronomy where
every telescope kicks in where (and only where) it is best needed.

\section{Future}

I strongly believe NOT will continue to be competitive due to its
size, location and flexibility. At Stockholm University, half a dozen
of scientists have joined the Vera Rubin Observatory to conduct the
Large Synoptic Survey Telescope (LSST) survey over the coming 10 years
(starting 2024). This will find tens of thousands of transients and
will again transform time domain astronomy.
We have offered to provide $\sim300$ hours of NOT
telescope time over three years to the LSST project for
classifications of supernovae,
and for this we, of course, hope to use the new Nordic Transient Explorer
(NTE) with its improved sensitivity and extended wavelength range.
There will be no lack of targets for future supernova
astronomers, the challenge will be to cope with the flood of data and
cleverly select the most interesting targets for follow-up.

\section*{Acknowledgements}

Based on observations made with the Nordic Optical Telescope, owned
in collaboration by the University of Turku and Aarhus University, and
operated jointly by Aarhus University, the University of Turku and the
University of Oslo, representing Denmark, Finland and Norway, the
University of Iceland and Stockholm University at the Observatorio del
Roque de los Muchachos, La Palma, Spain, of the Instituto de
Astrofisica de Canarias.
The data presented here were obtained [in part] with ALFOSC, which is
provided by the Instituto de Astrofisica de Andalucia (IAA) under
a joint agreement with the University of Copenhagen and NOT.
In my talk I ended with a slide saying: {\bf Thanks a lot NOT!} It has
been a real pleasure. Send any comments on this to jesper@astro.su.se.
Jesper Sollerman is professor at the Department of Astronomy at
Stockholm University and affiliated to the Oskar Klein Centre. He was
the Swedish representative of NOTSA and is the Stockholm
representative in the new NOT board. He has regularly
been using the NOT since 1994.

\bibliography{notref}
  
\end{document}